\documentclass{article}

\usepackage{graphicx}
\usepackage[space]{grffile}
\usepackage{latexsym}
\usepackage{textcomp}
\usepackage{longtable}
\usepackage{tabulary}
\usepackage{booktabs,array,multirow}
\usepackage{amsfonts,amsmath,amssymb,bm}
\usepackage[round,sort&compress]{natbib}
\usepackage{url}
\usepackage{hyperref}
\hypersetup{colorlinks=false,pdfborder={0 0 0}}
\usepackage{etoolbox}
\makeatletter
\patchcmd\@combinedblfloats{\box\@outputbox}{\unvbox\@outputbox}{}{%
  \errmessage{\noexpand\@combinedblfloats could not be patched}%
}%
\makeatother

\AtBeginDocument{\DeclareGraphicsExtensions{.pdf,.PDF,.eps,.EPS,.png,.PNG,.tif,.TIF,.jpg,.JPG,.jpeg,.JPEG}}

\usepackage[utf8]{inputenc}
\usepackage[english]{babel}

\usepackage{siunitx}

\ifpdf
\hypersetup{
  pdftitle={AnEnDA with VAEs},
  pdfauthor={I. Grooms}
}
\fi

\begin{document}
\title{Analog ensemble data assimilation and a method for constructing analogs with variational autoencoders}

\author{Ian Grooms}

\date{Department of Applied Mathematics, University of Colorado, Boulder}
\maketitle
\selectlanguage{english}
\begin{abstract}
It is proposed to use analogs of the forecast mean to generate an ensemble of perturbations for use in ensemble optimal interpolation (EnOI) or ensemble variational (EnVar) methods.
A new method of constructing analogs using variational autoencoders (VAEs; a machine learning method) is proposed.
The resulting analog methods using analogs from a catalog (AnEnOI), and using constructed analogs (cAnEnOI), are tested in the context of a multiscale Lorenz-`96 model, with standard EnOI and an ensemble square root filter for comparison.
The use of analogs from a modestly-sized catalog is shown to improve the performance of EnOI, with limited marginal improvements resulting from increases in the catalog size.
The method using constructed analogs (cAnEnOI) is found to perform as well as a full ensemble square root filter, and to be robust over a wide range of tuning parameters.
\end{abstract}
\section{Introduction}
Data assimilation methods are widely used in geophysics for a variety of purposes. 
Workhorse methods include the Ensemble Kalman Filter (EnKF) and its many variants \citep{Evensen94,HM98,BvLE98}, and 3D-Var and 4D-Var \citep{Talagrand10}.
Traditional variational methods suffer from the use of a time-independent background covariance, whereas the drawbacks of the EnKF include the sometimes high cost of generating ensemble members and less accurate treatment of nonlinearity and non-Gaussianity.
A variety of hybrids exist between ensemble and variational methods that aim to combine the strengths of the different methods \citep{Bannister17}.
Ensemble optimal interpolation \citep[EnOI;][]{OAMEK02,Evensen03} uses a time-independent background covariance that is generated from a time-independent ensemble of perturbations.
In EnOI a single model simulation is required for each assimilation cycle to propagate the mean state.
EnOI uses a gain matrix to compute the increment between the forecast and analysis means.
One could alternatively use a variational approach that finds the minimizer of a loss function whose background covariance is defined using the time-independent ensemble of perturbations; such a method would be a form of EnVar \citep{Lorenc13}.

The ensemble of perturbations used in EnOI usually comes from a catalog of model states from a long-running simulation.
Since the ensemble of perturbations used in EnOI is static EnOI suffers from the same drawbacks as classical variational methods, but has the benefit that only a single forecast is required for each assimilation cycle.
The goal of this investigation is to explore a way of improving the performance of EnOI by generating a time-dependent ensemble of perturbations from a large catalog.
The premise is that an ensemble of model states chosen as a subset from the catalog that are similar to the current forecast will produce an ensemble of perturbations that is more appropriate for use in EnOI than an ensemble that is representative of the climatology of the model.
Ensemble perturbations drawn from the climatology represent the correlations in the climatology, which can be a poor proxy for correlations in the forecast error.
Analog ensemble perturbations come from the part of the dynamical system's attractor (or pullback attractor for non-autonomous systems) that is close to the actual forecast, and therefore represent correlations on a specific part of the model attractor rather than over the whole climatology.
As a result they are expected to provide a more realistic representation of forecast error, since the forecast error distribution should be expected to cover a neighborhood of the attractor close to the forecast mean. 

Model states that are similar to the current forecast are called `analogs' \citep{Lorenz69} and have a long history in weather forecasting and forecast downscaling \citep{DERNS13,ED16,ZG16}.
\citet{vdD95} considered finding analogs from a large historical catalog of model states, and showed that to make an effective analog global weather forecast would require an impossibly large catalog - on the order of $10^{30}$ years of data.
Nevertheless, in the current setting one may still expect some degree of success with analogs drawn from a practically-sized catalog since the analogs are not being used for forecasting, but only to improve the background covariance within the data assimilation framework.

One way of avoiding the impossibly large size requirements of a catalog for analog forecasting is to use a reasonably-sized catalog to construct analogs, and there are many ways of doing this \citep{HDC08,MHDDC10,AB12,PCT14}.
This investigation explores a new way of constructing analogs using variational autoencoders \citep{KW19}.
A standard autoencoder consists of two functions: an encoder $\bm{e}(\bm{x})$ that maps the model state $\bm{x}\in\mathbb{R}^d$ to a latent space $\bm{z}\in\mathbb{R}^l$ where $l\ll d$, and a decoder $\bm{d}(\bm{z})$ that maps a vector in the latent space to a model state.
Both $\bm{e}$ and $\bm{d}$ are usually specified as deep artificial neural networks.
Given a catalog of model states $\{\bm{x}_i\}_{i=1}^N$, the parameters of $\bm{e}$ and $\bm{d}$ are chosen to minimize
\[\sum_i \|\bm{x}_i - \bm{d}(\bm{e}(\bm{x}_i))\|_2^2\]
or some similar loss function.
A standard autoencoder does not impose any particular structure on the latent space.
For example, a sufficiently powerful autoencoder might simply learn the map $i = \bm{e}(\bm{x}_i)$, $\bm{d}(i) = \bm{x}_i$.
As a result, standard autoencoders are not always useful as generative models: If $\bm{z}_i = \bm{e}(\bm{x}_i)$, then $\bm{d}(\bm{z}_i+\bm{\epsilon})$ need not be very similar to $\bm{x}_i$ for small $\bm{\epsilon}$.
Variational autoencoders attempt to impose structure in the latent space; specifically, they aim to choose the parameters of $\bm{e}$ and $\bm{d}$ so that the structure of the data in latent space is approximately Gaussian. 
This is accomplished by two devices.
First, the latent space is divided in two so that $\bm{e}(\bm{x}) = (\bm{\mu},\bm{\sigma})$ where $\bm{\mu},\bm{\sigma}\in\mathbb{R}^l$.
Then, a latent space vector is constructed as $\bm{z} = \bm{\mu} + \bm{\sigma}\circ\bm{\epsilon}$ where $\bm{\epsilon}$ is a standard normal Gaussian random vector and $\circ$ denotes the elementwise product (also known as Hadamard product, or Schur product).
Second, the loss function is altered by the addition of a term that penalizes deviations of the latent space distribution from a standard normal.
(For details on the form of this additional penalty term see \citet{KW19}.)
This investigation uses a variational autoencoder (VAE) to generate analogs that are then used to construct the ensemble of perturbations for use in data assimilation.

Data assimilation algorithms using analogs have been proposed in the context of geophysical data assimilation by \citet{LTAPF17,LguensatEtAl19}.
Two key differences between this approach and the current approach are (i) the current approach relies on a model simulation to make the forecast, whereas in \citet{LTAPF17,LguensatEtAl19} the dynamics are data-driven in a manner similar to analog forecasting, and (ii) the current approach investigates the use of VAEs to construct analogs.
The data-driven approach of \citet{LTAPF17,LguensatEtAl19} is expected to be clearly superior in cases where there is no reliable dynamical model for the system in question, or where such a model would be prohibitively expensive.

The investigation is carried out in the context of a multiscale Lorenz-`96 model, which is described in section \ref{sec:L96}.
The configuration and training of the VAE is described in section \ref{sec:VAE}.
The data assimilation system setup is described in section \ref{sec:DA}, and the results of data assimilation experiments are described in section \ref{sec:Results}.
Conclusions are offered in section \ref{sec:Conclusions}.

\section{Multiscale Lorenz-`96 model configuration}
\label{sec:L96}
Many data assimilation methods have been initially explored in the context of the Lorenz-'96 model \citep{Lorenz96,Lorenz06}.
Higher dimensionality can be obtained in this model by simply retaining the model form while increasing the dimension; alternatively there is a two-scale version also described by \citet{Lorenz96}.
This latter two-scale model has two sets of variables, $X_i$ describing the large, slow scales and $Y_j$ describing the small, fast scales.
\citet{GL15} introduced a multiscale Lorenz-`96 model with a single set of variables $x_i$ with distinct large-scale and small-scale parts.
The model is governed by the following system of ordinary differential equations
\begin{equation}
\dot{\bm{x}} = h\bm{N}_S(\bm{x}) + J\mathbf{T}^T\bm{N}_L(\mathbf{T}\bm{x}) -\bm{x} + F\bm{1}
\end{equation}
where $h,F\in\mathbb{R}$, $J\in\mathbb{N}$, $\bm{1}$ is a vector of ones, and the nonlinearities have the form
\begin{align}
\left(\bm{N}_S(\bm{x})\right)_i &= -x_{i+1}(x_{i+2}-x_{i-1})\\
\left(\bm{N}_L(\bm{X})\right)_k &= -X_{k-1}(X_{k-2}-X_{k+1}).
\end{align}
The experiments presented here use $h=0.5$ and $F=8$.
The matrix $\mathbf{T}$ projects onto the 41 largest-scale discrete Fourier modes and then evaluates that projection at 41 equally-spaced points.
The matrix $J\mathbf{T}^T$ spectrally interpolates a vector of length 41 back to the full dimension of $\bm{x}$.
The number of state variables in $\bm{x}$ is $41J$; here $J=64$ for a total system dimension of 2624.
In the definition of the nonlinear terms the indices are assumed to extend periodically, as in the Lorenz-`96 model.

\begin{figure*}
    \centering
    \includegraphics[width=\textwidth]{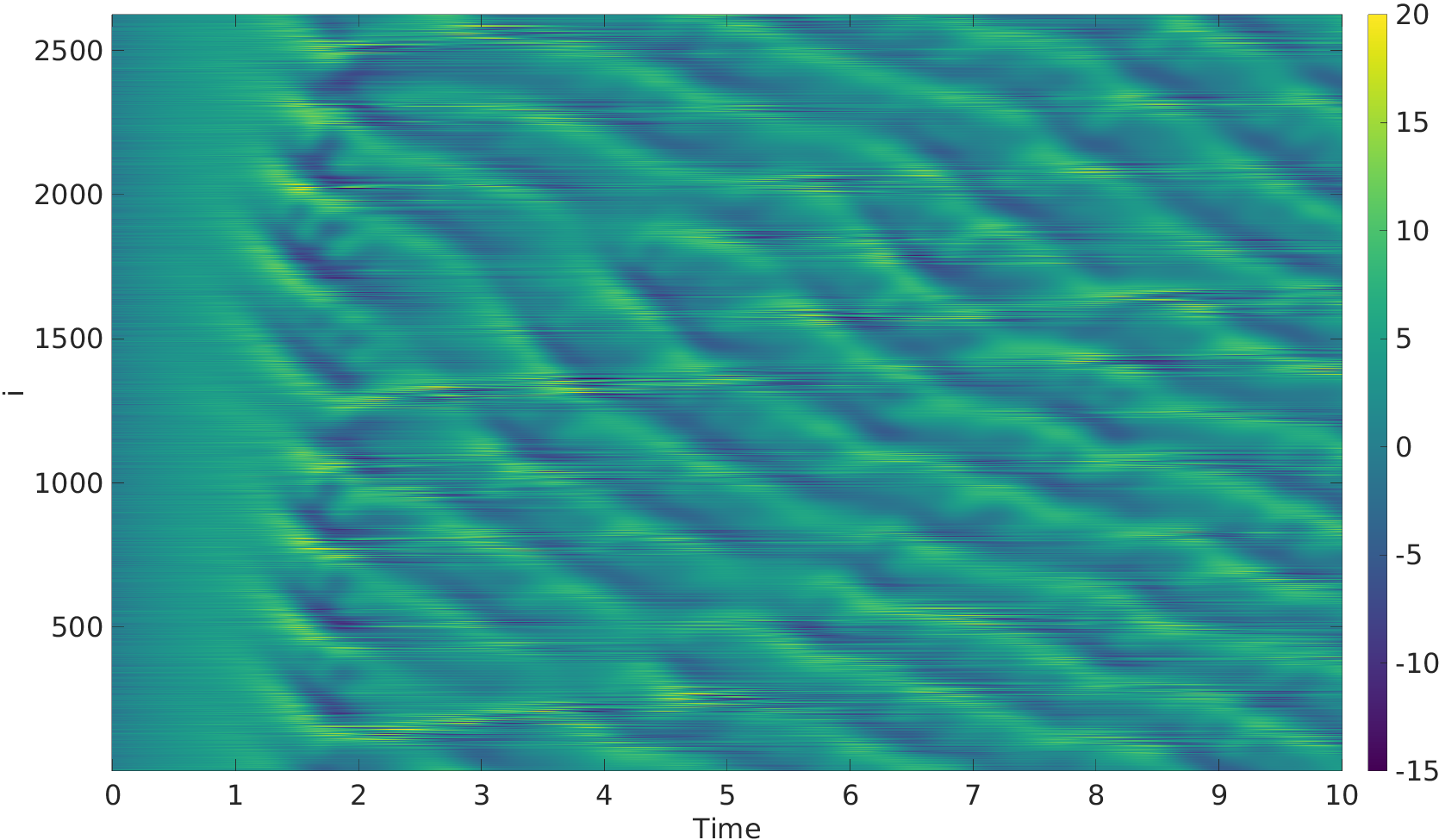}
    \caption{A simulation of the multiscale Lorenz-`96 model initialized at $t=0$ with a sample from a standard normal distribution.}
    \label{fig:L96}
\end{figure*}

The large-scale part of the model dynamics, which can be extracted by applying $\mathbf{T}$ to $\bm{x}$, is identical to the dynamics of the standard Lorenz-`96 model, except that the large scales are coupled to small scales via the term $h\mathbf{T}\bm{N}_S(\bm{x})$.
While the Lorenz-`96 model is often configured with $K=40$ large-scale variables \citep[e.g.][]{LE98}, this multiscale model uses 41 variables so that the real and imaginary parts of the 20$^\text{th}$ Fourier mode are not split between large and small scales.
At small scales, the dynamics are the same as those of original Lorenz-`96 model but with the direction of indexing reversed, and with coupling to the large scales.
Coupling to the large scales drives small-scale instabilities, which then grow and cause feedback onto the large-scale flow.
Figure \ref{fig:L96} shows the result of a simulation of this model initialized at $t=0$ with a sample from a standard normal distribution.
After a short transient the dynamics settle onto an attractor, with large-scale nonlinear waves propagating eastward and small-scale instabilities transiently excited by the large-scale waves.

\section{Variational Autoencoder}
\label{sec:VAE}
\begin{figure*}
    \centering
    \includegraphics[width=\textwidth]{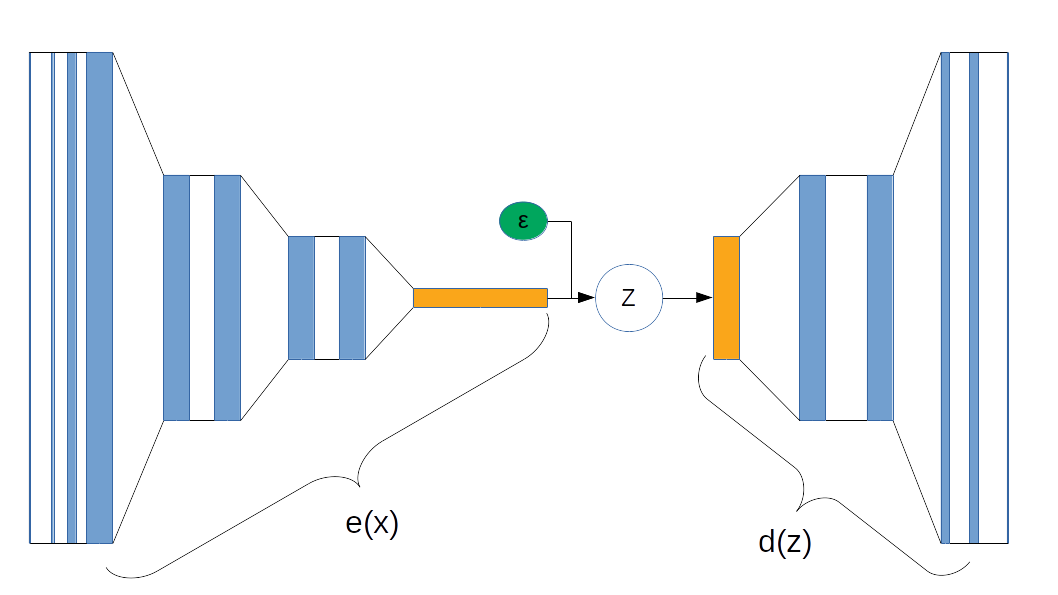}
    \caption{Architecture of the variational autoencoder. The leftmost vertical line indicates the data $\bm{x}$. The blue rectangles in the left half indicate convolutional layers, and the blue rectangles in the right half indicate transposed convolutional layers. The yellow rectangles in the middle indicate fully-connected layers. The green oval indicates the random noise $\epsilon$. The rightmost vertical line indicates the output.}
    \label{fig:VAE}
\end{figure*}

Variational autoencoders (VAEs) were described generally in the introduction; the architecture of the VAE used here is described in this section.
\citet{HH19} provide an introduction to machine learning with artificial neural networks and the associated terminology.
The architecture of the autoencoder is summarized in Fig.~\ref{fig:VAE}.
The encoder $\bm{e}(\bm{x})$ is constructed as follows:
\begin{enumerate}
\item A convolutional layer with three filters of size $3\times 1$
\item A convolutional layer with nine filters of size $3\times 1$
\item A convolutional layer with 27 filters of size $3\times 1$
\item A max pooling layer with $2\times1$ pool size
\item Two convolutional layers with 27 filters each of size $3\times 1$
\item A max pooling layer with $2\times1$ pool size
\item Two convolutional layers with 27 filters each of size $3\times 1$
\item A max pooling layer with $2\times1$ pool size
\item A fully connected layer with two outputs, each of size 492. 
\end{enumerate}

The decoder $\bm{d}(\bm{z})$ is constructed as follows:
\begin{enumerate}
\item A fully connected layer with output of size $656$ with 27 channels. 
\item A transposed convolutional layer with 27 filters of size $3\times 1$ and stride of 2
\item A transposed convolutional layer with 27 filters of size $3\times 1$ and stride of 1
\item A transposed convolutional layer with nine filters of size $3\times 1$ and stride of 2
\item A transposed convolutional layer with nine filters of size $3\times 1$ and stride of 1
\item A transposed convolutional layer with one filter of size $3\times1$ and a stride of 1.
\end{enumerate}

The convolutional layers, transposed convolutional layers, and fully connected layers all use the exponential linear unit activation function, with the form
\begin{equation}
\text{eLU}(s) = \left\{\begin{array}{cl}s&s\ge0\\e^{-s}-1&s<0\end{array}\right.
\end{equation}
The eLU function has a continuous first derivative, which implies that the the decoder also has a continuous first derivative, since it is a composition of continuously-differentiable functions.
By contrast, the use of max pooling layers in the encoder implies that the encoder is continuous, but its first derivative is only piecewise continuous.

The data used to train the VAE consists of 70,000 snapshots of the state of the multiscale Lorenz-`96 model described in the previous section.
These snapshots are generated by initializing the model from a standard normal, then running the simulation until it reaches a statistical equilibrium, then taking data every 1 time unit, which corresponds to 5 days in the standard dimensionalization of the Lorenz-`96 model.
Using training data from the model's attractor means that the variational autoencoder is attempting to learn a map that transforms the stationary invariant measure on the system attractor to a Gaussian distribution in latent space.
The model is trained (i.e. the parameters of $\bm{e}$ and $\bm{d}$ are estimated) using stochastic gradient descent.
The batch size is 3500 snapshots, and the optimization was trained for 272 epochs, at which point the objective function had saturated.
In more realistic applications than the multiscale Lorenz-`96 model it would be of interest to explore multiple architectures and training regimens to investigate whether the VAE strikes a balance between being sufficiently expressive and being simple enough to train with limited data.
For the purpose here of demonstrating the proof of concept in a simple model, a single architecture suffices.

\section{Data Assimilation: Methods}
\label{sec:DA}
The observations are taken at every fourth point in space and at every 0.2 time units (which corresponds to 1 day in the standard dimensionalization of the Lorenz-`96 model).
At every assimilation cycle there are effectively 16 observations for each of the 41 large-scale Lorenz-`96 modes.
The observation errors are Gaussian with zero mean and variance 1/2. 
With this observing system and this forecast lead time the dynamics of the model remain only weakly non-Gaussian according to the taxonomy of \citet{MCSB14} and \citet{MH19}, implying that the primary limitation to performance of an EnKF will be ensemble size rather than non-Gaussianity.

Each of the various data assimilation methods described below has several tunable parameters, e.g. localization radius and inflation factor.
To optimize the performance of each method, a range of parameters is explored.
For each parameter combination that is tested, at least 8 experiments are run.
For each experiment a reference simulation is initialized from standard normal noise and run for 9 time units, by which time it has reached a statistical equilibrium.
Observations are taken starting at time $t=9$, every $0.2$ time units (1 day) for 73 time units (one year), which corresponds to 365 assimilation cycles.
The first 73 assimilation cycles of each experiment are considered a burn-in period, and are discarded when computing performance statistics.

At each assimilation cycle the performance of the filter is measured using the root mean square error (RMSE), defined as the 2-norm of the error between the reference simulation and the filter analysis mean.
At each parameter value this procedure results in at least $8\times(365-73)=2336$ values of RMSE.
The mean of these values is used to summarize the performance of the method for that specific combination of parameters.
RMSE based on the forecast is available, but does not behave qualitatively differently from analysis RMSE and is therefore not shown.

The following subsections detail the different assimilation methods to be compared.

\subsection{Serial Ensemble Square Root Filter}
The point of this investigation is to consider methods that improve on EnOI but that are less computationally costly than an EnKF.
As such, it is useful to run an EnKF as a baseline for comparison.
The baseline method used here is the serial ensemble square root assimilation of \citet{WH02}, with Schur-product localization in observation space and multiplicative inflation.
This method is referred to as ESRF in the results.

The initial ensemble is constructed by initializing each ensemble member using an independent draw from a standard normal distribution, then forecasting this initial condition for 9 time units (45 days), by which time they have reached the system attractor.
The final condition of each simulation at $t=9$ is used to initialize the ensemble, so the initial ensemble is completely independent of the reference simulation used to generate the observations.
The localization function has the form
\begin{equation}
\ell_i = e^{-\frac{1}{2}\left(\frac{i}{L}\right)^2}
\end{equation}
where $L$ is the localization radius.
For reference, the large-scale Lorenz-`96 modes in this model are effectively 64 units apart, so to convert $L$ to a comparable localization radius for the standard Lorenz-`96 model it suffices to divide $L$ by 64.
The multiplicative inflation is applied to the analysis ensemble, since \citet{GRAW19} recently found that posterior inflation is more appropriate and more effective in situations without model error.
Inflation is applied by multiplying the analysis ensemble perturbations by an inflation factor of $r\ge 1$.

The three tunable parameters for the ESRF are the ensemble size $N_e$, the localization radius $L$, and the inflation factor $r$.
Some limited exploration of ensemble size $N_e$ was performed. 
First a range of $L$ and $r$ were explored at $N_e=100$.
Then, a range was explored at $N_e=200$.
The optimal RMSE obtained at $N_e=200$ was not significantly better than at $N_e=100$, so all results reported here for all methods (including ESRF) use an ensemble size of $N_e=100$.

\subsection{Ensemble OI}
EnOI can also be considered as a baseline for comparison of the analog methods, but on the other side from the ESRF method.
The EnOI used here is configured exactly the same as the ESRF, except that no inflation needs to be applied.
A different ensemble of perturbations is drawn randomly for each experiment from a catalog of 41,000 model states (once drawn, the ensemble perturbations remain time-independent for all assimilation cycles within a single experiment).
This catalog is different from the one used to train the VAE, but is constructed in the same way.
The climatological spread represented by this ensemble is too large to be an accurate representation of the forecast error, so the ensemble of perturbations is scaled to a pre-defined forecast spread, which forms the second tunable parameter (together with localization radius) for the EnOI method.

\subsection{Analog Ensemble OI}
The analog ensemble OI (AnEnOI) method is exactly the same as the EnOI method except for the following: At each assimilation cycle the ensemble is chosen to be the $N_e=100$ members of the catalog that are closest to the current forecast.
The impact of the size of the catalog is briefly explored by performing experiments using (i) a catalog of only 1,000 members, and (ii) the full catalog of 41,000 members.
Results reported below are for the smaller catalog, unless noted otherwise.
The similarity of analogs to the forecast is defined using the 2-norm; the impact of using other, more dynamically motivated measures of similarity is not explored.

\subsection{Constructed Analog Ensemble OI}
The constructed analog ensemble OI (cAnEnOI) is exactly the same as the AnEnOI except for the construction of the analogs.
To construct analogs, the forecast mean is first encoded using the encoder $\bm{e}(\bm{x})$.
Recall that the encoder produces two vectors, $\bm{\mu}$ and $\bm{\sigma}$, and during training the encoded state is $\bm{z} = \bm{\mu} + \bm{\sigma}\circ\bm{\epsilon}$ where $\bm{\epsilon}$ is a standard normal random variable.
For the purposes of constructing analogs, an ensemble in latent space is constructed as follows:
\begin{equation}
\bm{z}_i = \bm{\mu} + r_z \bm{\epsilon}_i,\;\;i=1,\ldots,N_e
\end{equation}
where $\bm{\epsilon}_i$ are independent draws from a standard normal distribution and $r_z$ is a tunable parameter governing the spread of the ensemble in the latent space.
The analog ensemble is then constructed using the decoder as $\bm{x}_i = \bm{d}(\bm{z}_i)$.

As noted above, the decoder is a continuously-differentiable function.
The ensemble in latent space is Gaussian, so for small enough $r_z$ the analog ensemble will also be approximately Gaussian distributed, with a covariance matrix approximately
\begin{equation}
r_z^2 \mathbf{DD}^T
\end{equation}
where $\mathbf{D}$ is the Jacobian derivative of $\bm{d}$ evaluated at $\bm{\mu}$.
The rank of this covariance matrix is less than or equal to the dimension of the latent space, since $\mathbf{D}$ is a $d\times l$ matrix.
Of course the analog ensemble covariance matrix will also have rank less than or equal to $N_e-1$.

For small $r_z$ the correlation structure depends only on the forecast mean, and not on $r_z$.
For larger $r_z$ the nonlinearity of the decoder comes into play with two important consequences.
First, the analog ensemble becomes increasingly non-Gaussian, which allows the rank of the covariance matrix to exceed the dimension of the latent space (though the ensemble covariance matrix still must have rank bounded by $N_e-1$).
Second, the correlation structure of the analog ensemble begins to depend on $r_z$ as well as on the forecast mean.

It is desirable to decouple the forecast spread of the analog ensemble from the correlation structure of the analog ensemble covariance matrix.
This can be achieved by first constructing the analog ensemble as described above, and then rescaling the ensemble perturbations to have the desired spread.
As a result, the cAnEnOI method has three main tunable parameters: (i) localization radius, (ii) $r_z$ which controls the correlation structure of the analog ensemble perturbation covariance matrix, and (iii) the forecast spread.

\section{Data Assimilation: Results}
\label{sec:Results}
\begin{figure*}
    \centering
    \includegraphics[width=\textwidth]{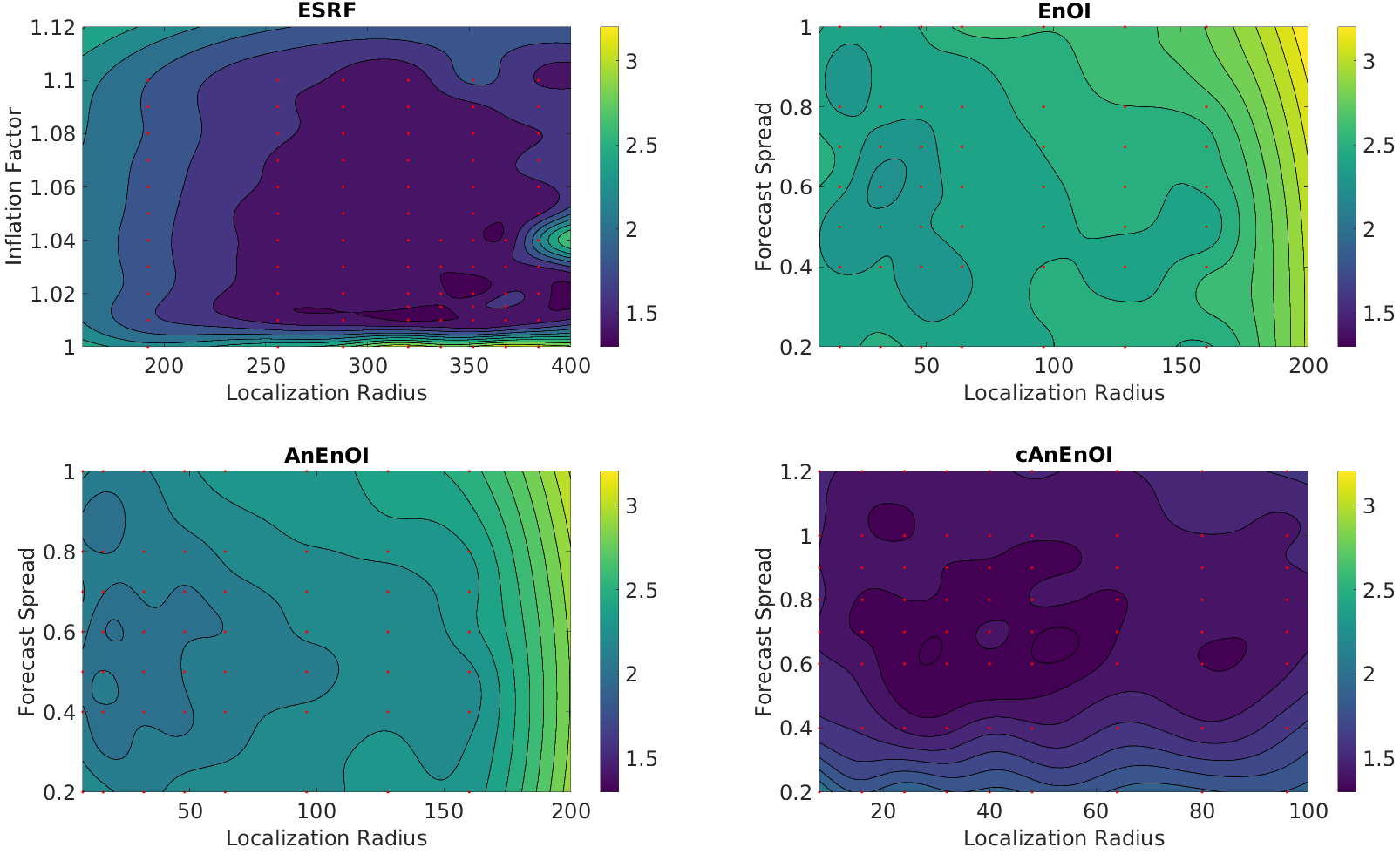}
    \caption{Mean RMSE for the four methods, as a function of the governing parameters. The experimental results are shown as red dots; values in between are interpolated. Values for cAnEnOI are at $r_z=0.7$. The colorbar is the same for all plots. Note that all methods include localization radius as a tunable parameter, but ESRF has inflation as a tunable parameter while the other methods have forecast spread. The axis limits on each panel are different.}
    \label{fig:RMSE}
\end{figure*}
Figure \ref{fig:RMSE} shows analysis RMSE as a function of tunable parameters for the four methods (ESRF, EnOI, AnEnOI, and cAnEnOI).
Results for the cAnEnOI method are shown for $r_z=0.7$; the dependence on $r_z$ is discussed below.
The ESRF has a fairly broad well in parameter space where the analysis RMSE is around 1.5.
The optimal observed RMSE is 1.35, which occurs at inflation factor $r=1.02$ and localization radius $L=320$.
This is about a factor of two larger than the raw observation error $1/\sqrt{2}$, which is not bad given that only one quarter of the state variables are observed.
At localization radii smaller than 192 or larger than 384 the performance begins to degrade.
For larger localization radii the ESRF performance becomes erratic, being limited by the deleterious effects of rare spurious long-range correlations: some experiments perform well, while others diverge.
For smaller localization radii the ESRF performance also degrades, for reasons that are not entirely clear and appear to be related to dynamical imbalance of the analysis.

The EnOI (with a catalog of 1,000 model states) has an optimal analysis RMSE of 2.27, which occurs at a localization radius of 32 and a forecast spread of 0.6.
Though significantly worse than ESRF, the EnOI still produces reasonably-accurate analyses; for comparison, a random draw from the climatological distribution would produce an RMSE of 4.97.
The optimal localization radius for EnOI is a factor of 10 smaller than for ESRF.
This is presumably because the correlations encoded in the ESRF ensemble are far more meaningful (i.e. representative of forecast error correlations) at long range than the climatological correlations associated with the EnOI ensemble.

The use of analogs significantly improves the EnOI method: the optimal analysis RMSE for AnEnOI is 2.01, which occurs at a localization radius of 16 and a forecast spread of 0.6.
It is not clear why the optimal localization radius decreases, but on the other hand as seen in Fig.~\ref{fig:RMSE} the analysis RMSE of AnEnOI is not too strongly sensitive to changes in localization radius or forecast spread.
This is very encouraging, since a catalog of only 1,000 states would presumably be far too small to produce accurate analog forecasts for this system.
Increasing the catalog size to 41,000 further improves the analysis RMSE to 1.90: a very modest improvement for a very large increase in catalog size.
To put a positive spin on this, it suggests that the bulk of the benefits that can be obtained by moving from EnOI to AnEnOI do not require an unrealistically large catalog.

The real success comes from using constructed analogs.
The optimal analysis RMSE obtained using cAnEnOI at $r_z=0.6$ is 1.30: slightly better than obtained using ESRF!
The optimal localization radius and forecast spread are 40 and 0.7, respectively, but as shown in Fig.~\ref{fig:RMSE}, the performance of cAnEnOI is not strongly sensitive to changes in these parameters: performance comparable to ESRF can be obtained over a wide range of localization radii and forecast spread.

\begin{figure*}
    \centering
    \includegraphics[width=\textwidth]{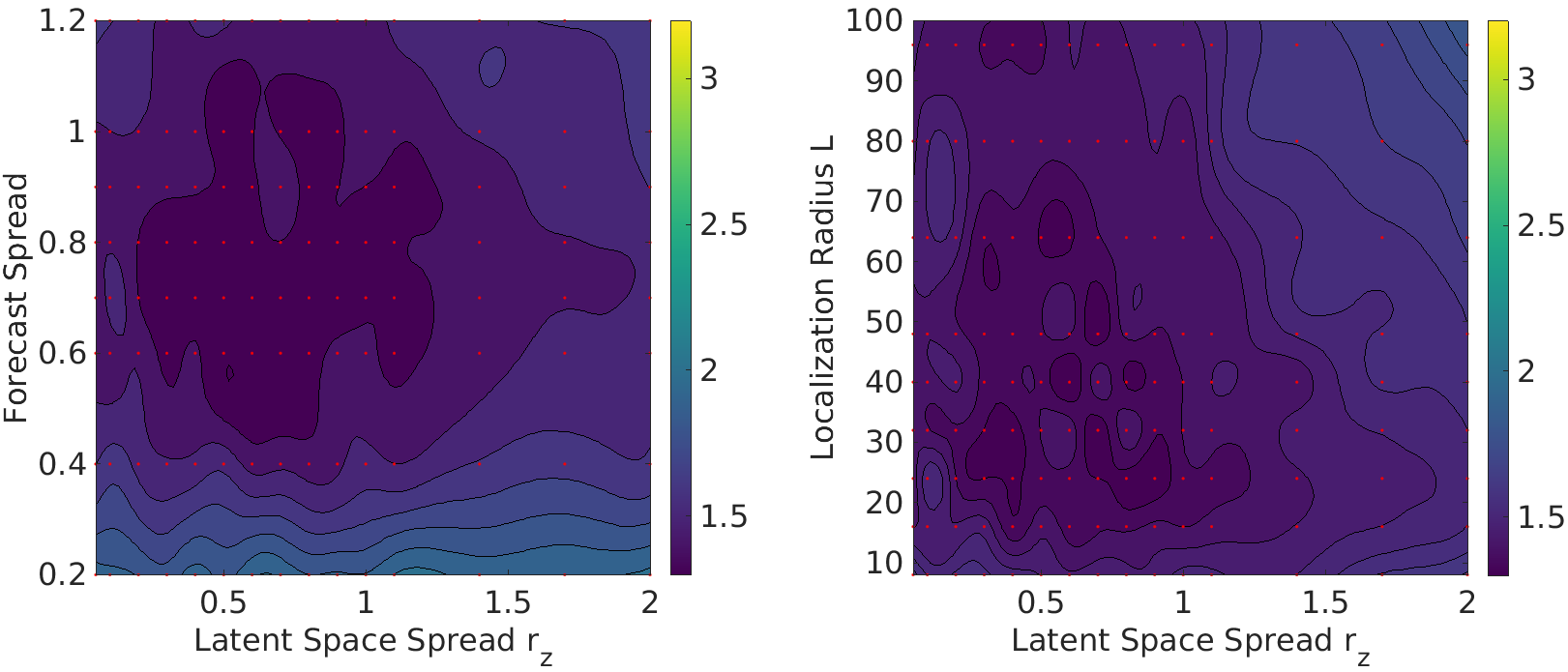}
    \caption{Mean RMSE for the cAnEnOI method, as a function of latent space spread $r_z$ and forecast spread (left panel), and as a function of $r_z$ and localization radius $L$ (right panel). The experimental results are shown as red dots; values in between are interpolated. The colorbar is the same for both panels, and is the same as in Fig.~\ref{fig:RMSE}.}
    \label{fig:RZ}
\end{figure*}

Figure \ref{fig:RZ} shows the cAnEnOI analysis RMSE as a function of latent space spread $r_z$ for a fixed localization radius of 24 (left panel) and a fixed forecast spread of 0.7 (right panel).
The method is extremely robust to varying all three parameters (forecast spread, localization radius, and latent space spread), and is able to produce RMSE comparable to ESRF over a wide range of parameters.
As noted above, the correlation structure of the constructed analog ensemble is independent of $r_z$ for small $r_z$.
Consistent with this, for $r_z$ between $0.05$ and $0.2$, cAnEnOI produces RMSE of 1.38 (at optimal values of forecast spread and localization radius), which is comparable to ESRF.
As $r_z$ increases the performance improves, with excellent results in the range $.2\le r_z\le 1$.
As $r_z$ increases further the performance slowly degrades, but even at $r_z=2$ the performance is better than the optimal results using the AnEnOI method with `found' analogs.

\section{Conclusions}
\label{sec:Conclusions}
This work introduces a new use for analogs, besides forecasting and downscaling: to construct an ensemble background covariance matrix for use in data assimilation, as in the EnOI or EnVar frameworks.
The research was carried out in the context of a multiscale Lorenz-`96 model invented by \citet{GL15}.
Two methods were formulated: one based on finding analogs within a catalog of historical states (AnEnOI), the other based on constructing analogs using a variational autoencoder \citep[VAE;][]{KW19} trained on a catalog of historical states (cAnEnOI).
It was found that AnEnOI outperforms a basic EnOI method even with a relatively small catalog of 1,000 members, and further improvements were marginal when the catalog size was increased to 41,000.
The cAnEnOI method was able to perform as well as an optimized ensemble square root filter (ESRF), and was quite robust to variations in the tuning parameters of the method.
Several alternate methods exist for constructing analogs \citep{HDC08,MHDDC10,AB12,PCT14}; these could also be used in a cAnEnOI method.

In real geophysical applications the model states are much larger than in the simple model considered here.
This leads to two difficulties in implementing the analog ensemble data assimilation methods proposed here: (i) finding analogs within the catalog is expensive, and (ii) training a VAE to reproduce an entire model state is likely far more difficult and may be practically impossible.
Fortunately, given the long history of analogs, there is already research on efficient ways to find analogs within a large catalog of large model states; see, e.g., \citet{RDPL18} and \citet{YA19}.
Since the method using constructed analogs is far more successful, the second difficulty of real geophysical models is more pertinent.
To overcome this limitation it is suggested to use a local analysis in the vein of the Local Ensemble Kalman Filter \citep[LEnKF;][]{BrusdalEtAl03,Evensen03,OttEtAl05}.
This framework uses many local ensembles: for each model grid point a local ensemble analysis is performed using observations near that grid point.
The cAnEnOI method developed here could easily be used in this local framework: For each grid point an analog ensemble is constructed for use in the local assimilation.
The benefit of such a local analysis is that the VAE would only have to be trained to generate local subsets of the model state, rather than, e.g., the full state of a global coupled climate model.
A similar localization procedure could be leveraged in the case of `found' analogs rather than constructed ones.

Overall, the results are quite promising.
EnOI is a widely used method \citep{XCZBS11,BCJP14,MTSLX15,DLQZZ18,WZZ18} because of its acceptable performance and significantly reduced cost compared to EnKF, and ensemble background covariances are widely used in EnVar and hybrid data assimilation methods \citep{Bannister17}.
The results here suggest that improvements could be obtained using either found analogs or constructed analogs; the increased cost of using analogs will be situation-dependent, but if the costs can be made lower than the cost of forecasting an ensemble, then the analog EnOI or EnVar methods may be an attractive alternative.

\section*{Acknowledgments}
The author is grateful to Jeff Anderson for a discussion on the history of analog weather forecasting.
This work used the Extreme Science and Engineering Discovery Environment \citep[XSEDE;][]{XSEDE} Bridges \citep{Bridges} at the Pittsburgh Supercomputing Center through allocation TG-DMS190025.
This work was funded by the US National Science Foundation under grant number DMS 1821074.

\end{document}